\documentclass[a4paper,11pt]{article}
\usepackage[affil-it]{authblk} 
\usepackage{amsmath}
\usepackage{amsfonts}
\usepackage{txfonts}
\usepackage{graphicx}
\usepackage{dcolumn}
\usepackage{bbm}
\usepackage{amssymb}
\usepackage{latexsym}
\usepackage{titlesec}
\usepackage[colorlinks=true, linkcolor=red, citecolor=blue]{hyperref}
\usepackage[numbers,sort&compress]{natbib}
\usepackage{geometry}

\begin{document}
\title{\boldmath Probing the Neutrino Mass Hierarchy beyond $\Lambda$CDM Model}


\author[a]{En-Kun Li\thanks{ekli\_091@mail.dlut.edu.cn}}
\author[a]{Hongchao Zhang}
\author[a]{Minghui Du}
\author[a]{Zhi-Huan Zhou}
\author[a]{Lixin Xu\thanks{Corresponding author: lxxu@dlut.edu.cn}}

\affil[a]{Institute of Theoretical Physics, School of Physics, Dalian University of Technology, \\ Dalian, 116024, China}

\renewcommand\Authands{, and }

\maketitle

\begin{abstract}
    Taking the neutrino oscillation data into consideration, a dimensionless parameter $\Delta = (m_3-m_1)/(m_3+m_1)$ is adopted to parameterize the three neutrino mass eigenstates and the normal (positive $\Delta$) or inverted (negative $\Delta$) mass hierarchies in three typical cosmological models.
    Using the currently available cosmic observational data,  several Markov Chain Monte Carlo chains are obtained with uniform priors on the free parameters at first.
    Applying importance sampling the results are compared with three new priors, i.e., logarithmic prior on $|\Delta|$, linear and logarithmic priors on $\Sigma m_\nu$.
    It turns out that the three new priors increase the upper limits of neutrino mass, but do not change the tendency towards different model's preference for different hierarchies, i.e., the normal hierarchy tends to be favored by $\Lambda$CDM and $w$CDM, which, however, disappears in the $w_0 w_a$CDM model.
    In addition, the almost symmetrical contours in the $w-\Delta$, $w_0-\Delta$, $w_a-\Delta$ planes indicate that the normal and inverted hierarchy have strong degeneracy.
    Finally, we perform a Bayesian model comparison analysis, finding that flat linear prior on $\Delta$ and $w_0 w_a$CDM are the most preferred prior and model, respectively.
\end{abstract}

\flushbottom

\section{Introduction}
\label{sec:intro}

    The discovery of neutrino oscillations suggests that at least two of the three neutrino mass eigenstates are non-zero.
    And the current neutrino oscillation experiments can only give the squared differences between mass eigenstates \cite{Fukuda:1998mi, Lesgourgues:2006nd, GonzalezGarcia:2007ib, lesgourgues2013neutrino}.
    In the minimal three neutrino scenario, one of the squared difference is determined by solar oscillation experiments \cite{Araki:2004mb,Abe:2008aa} (i.e. $\Delta m_{21}^2 = m_2^2 - m_1^2$), and the other is measured by atmospheric oscillations \cite{Ashie:2005ik} (i.e. $|\Delta m_{31}^2| = |m_3^2 - m_1^2|$).
    Since the sign of $\Delta m_{31}^2$ remains undetermined, we are left with two logical possible mass hierarchies: the normal hierarchy (NH, $m_1 < m_2 < m_3$) and the inverted hierarchy (IH, $m_3< m_1 <m_2$).
    Moreover, the degeneracy of the three masses (i.e., $m_1 \sim m_2 \sim m_3 \gg |\Delta m_{31}|$) cannot be ruled out by further measurements of absolute mass \cite{deGouvea:2013onf}.
    Future terrestrial experiments, such as exploring the matter effects on earth with long baseline accelerators \cite{Acciarri:2015uup} and atmospheric neutrino experiments \cite{Aartsen:2014oha, Adrian-Martinez:2016fdl}, could shed light on the sign of $\Delta m_{31}^2$.

    Massive neutrinos are relativistic in the early universe and turn non-relativistic around the time of photon decoupling. 
    They can influence the early cosmological perturbations and leave characteristic imprints on the Cosmic Microwave Background (CMB) anisotropies \cite{Archidiacono:2016lnv} via early Integrated Sachs-Wolfe (ISW) effect.
    So the cosmological observations can play an import role in detecting the effects of the neutrino mass \cite{Hu:1997mj, Jimenez:2010ev}.
    Moreover, the early relativistic neutrinos suppress the clustering of matter and the late non-relativistic neutrinos contribute to matter cluster, so the growth of structure could be modified by neutrinos.
    Thus, one might extract useful signals of cosmic neutrinos from matter clustering and CMB \cite{Giusarma:2012ph,Giusarma:2016phn, Hannestad:2016fog,Vagnozzi:2017ovm, Simpson:2017qvj,Schwetz:2017fey, Capozzi:2017ipn,Caldwell:2017mqu}.
    Furthermore, the observations of the 21 cm line radiation and the CMB polarization \cite{Oyama:2015gma}, the cross-correlation between the Rees-Sciama effect and the weak lensing \cite{Xu:2016jns} are quite helpful to measure neutrino masses.
    More about the effects of neutrinos see e.g., \cite{Lesgourgues:2006nd} for a review.

     However, the constraint on the neutrino mass strongly depends on the nature of dark energy in the universe.
     Dark energy can affect the CMB power spectrum through altering the expansion rate of the universe and the gravitational potential (late ISW effect).
     Neutrinos, which behave as radiation at early times and contribute to matter density of late times, can also affect the CMB power spectrum similar (early ISW effect and late ISW effect) \cite{Hu:2001bc,Ichikawa:2004zi,Zhao:2016ecj}.
     Recently, a stringent upper bound of the total neutrino mass, i.e. $\sum m_\nu < 0.17$ eV ($95\% $) is provided by \textit{Planck} 2015 \cite{Ade:2015xua} using the data sets \textit{Planck} TT, TE, EE+lowP+BAO with the basic $\Lambda$CDM model.
     In Ref. \cite{Hannestad:2005gj}, S. Hannestad studied the influence of $w$CDM model with an arbitrary constant equation of state parameter (EOS) on the neutrino masses.
     The work suggests the existence of degeneracy between neutrino masses and the EOS of dark energy.
     Therefore, the constraint on the neutrino masses could be modified by a dynamical dark energy.
     Compared with the $\Lambda$CDM model, the constraint on the upper bound of the total neutrino mass is looser for the $w$CDM model, while tighter for the holographic dark energy model \cite{Zhang:2015uhk}.
     Moreover, previous studies have not yet taken the neutrino mass hierarchy into account, once it is considered, the cosmological data tends to favor the NH case in $\Lambda$CDM model \cite{Huang:2015wrx, Gerbino:2016ehw, Wang:2017htc}.
     Furthermore, the same conclusion still holds when the $w$CDM and holographic dark energy model are considered \cite{Wang:2016tsz}.
     On the other hand, different neutrino hierarchies mildly change the dynamical dark energy properties \cite{Yang:2017amu}.

     Taking the neutrino mass hierarchy into account, the cosmological parameters will be constrained by the publicly available Cosmological Monte Carlo (\textbf{CosmoMC}\footnote{ \url{http://cosmologist.info/cosmomc}} ) code \cite{Lewis:2002ah} in this paper.
     However, to study different mass hierarchies (NH, IH and even the degenerate case), we have to run the code at least twice, which requires a lot of computing resources.
    Here, we follow the method raised in Refs.\cite{Jimenez:2010ev,Xu:2016ddc}, especially in Ref. \cite{Xu:2016ddc}.
    In this paper, the authors introduced a hierarchy parameter to represent the two hierarchies, see Sec. \ref{sec:delta} for a brief review.
    Adopting the hierarchy parameter, one can save a lot of computing resources.

    Bayesian inference is a commonly used analytical method when cosmological observations are used to constrain the upper limits of $\sum m_\nu$ \cite{sivia1996data, Long:2017dru}.
    In this framework, the prior probability distribution of the total neutrino mass $\pi(\sum m_\nu)$ needs to be selected.
    Some recent studies have investigated the influences of the prior $\pi(\sum m_\nu)$ on cosmological parameters and the odds ratio of NH versus IH case \cite{Hannestad:2016fog, Gerbino:2016ehw, Giusarma:2016phn, Vagnozzi:2017ovm, Simpson:2017qvj, Schwetz:2017fey, Long:2017dru, Hannestad:2017ypp}.
    It is found that the constraints on $\sum m_\nu$ can change dramatically from different priors.
    In this paper, the neutrino masses are derived from the parameter $\Delta$, so its prior $\pi(\Delta)$ also has an effect on the constraints on $\sum m_\nu$.
    Therefore, we will use the importance sampling technique \cite{Lewis:2002ah} to investigate the influences of different $\pi(\Delta)$ on the neutrino masses and mass hierarchy.
    And the Bayesian evidence will be adopted to test the various cosmological models and prior choices.
    The outline of this paper is as follows.
    In Sec. \ref{sec:mod}, we give a brief introduction of the parametrized dark energy models and the method of formulating the neutrino mass hierarchy parameter.
    Section \ref{sec:datmet} describes the methodology and the cosmological data sets used in this paper.
    In Sec. \ref{sec:result}, we analyze the constraints of the neutrino masses and mass hierarchy in the $\Lambda$CDM, $w$CDM, and $w_0w_a$CDM models with different priors on $\Delta$.
    The paper concludes in Sec. \ref{sec:conclusion} where we summarize our results.

\section{Basic equations}
\label{sec:mod}

\subsection{Parametrized dark energy models}
\label{sec:de}

    Within the General Relativity framework, the line element for a spatially flat Friedmann-Lema\^{i}tre-Robertson-Walker (FLRW) universe is given by
    \begin{equation}
        ds^2 = -dt^2 + a^2(t) \left[ dr^2 + r^2 (d\theta^2 + \sin^2\theta d\varphi^2) \right],
    \end{equation}
    where $a(t)$ is the scale factor of the universe, and its present value is $a=a_0 =1$. 
    Suppose the universe is filled with photons ($r$), neutrinos ($\nu$), baryons ($b$), cold dark matter ($dm$), and dark energy ($de$) fluids.
    Now, in such a background, the Friedmann equations are
    \begin{align}
        & H^2 = \left( \frac{\dot{a}}{a}\right)^2 = \frac{8\pi G}{3} \sum_i \rho_i ,\\
        & \frac{\ddot{a}}{a} = -\frac{4\pi G}{3} \sum_i \left( \rho_i + 3 p_i\right)
    \end{align} 
    where $\cdot = d/dt$, $\rho_i$ and $p_i$ ($i = r, \nu, b, dm, de$) are the energy densities and the corresponding pressures of the $i^{th}$ component of the fluid, respectively.
    We also assume that there is no interaction between the fluids.
    Hence, the conservation equation of each component reads
    \begin{equation}
        \dot{\rho_i} + 3 H(\rho_i +p_i) = 0.
    \end{equation}
    
    If we consider a dynamic dark energy model with time-varying EOS, then its evolution equation will be
    \begin{equation}
        \rho_{de} = \frac{3H_0^2}{8\pi G} \Omega_{de} \exp{\left[ -3 \int_{a_0}^a \frac{1+w_{de}}{a'} da' \right]},
    \end{equation}
    where $H_0$ is the present Hubble parameter and $\Omega_{de}$ is the present dimensionless energy density of dark energy.
    Here we consider a typical parameterized dark energy model, i.e. the Chevallier-Polarski-Linder (CPL) model \cite{Chevallier:2000qy}, with the following energy density
    \begin{equation}
        \rho_{de} = \frac{3H_0^2}{8\pi G} \Omega_{de} a^{-3(1+w_0+w_a)} e^{-3w_a(1-a)},
    \end{equation}
    where $w_0$ is the present EOS of dark energy, and $w_a = dw/dz|_{z=0}$.
    Thus the cosmology model could be named as $w_0 w_a$CDM model. Note that when $w_a = 0$ this model reduces to $w$CDM and when $w_0=-1$, $w_a=0$ this model is the standard $\Lambda$CDM model.

\subsection{Neutrino mass hierarchy parameter}
\label{sec:delta}

\begin{figure}
    \centering
    \includegraphics[scale=0.5]{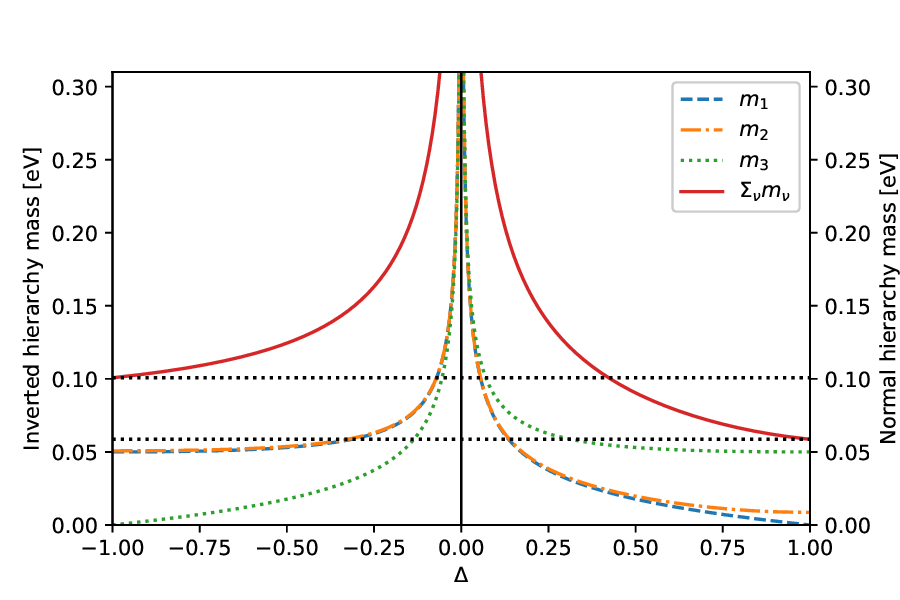}
    \caption{The masses of the three species of neutrinos and their total mass $\sum m_\nu$ with respect to the hierarchy parameter.}
    \label{fig:mass}
\end{figure}

    Taking the neutrino mass-squared differences into account and neglecting the experimental uncertainties, the two independent mass splitting parameters can be written as \cite{Agashe:2014kda}
    \begin{align}
        & \Delta m_{21}^2 \equiv m_2^2 -m_1^2 = 7.5 \times 10^{-5} {\rm eV}^2, \\
        & |\Delta m_{31}^2| \equiv |m_3^2 -m_1^2| = 2.5 \times 10^{-3} {\rm eV}^2.
    \end{align}
    Following the method raised in \cite{Xu:2016ddc}, the dimensionless mass hierarchy parameter in the range of $[-1,1]$ is defined as
    \begin{equation}
        \Delta = \frac{m_3-m_1}{m_3+m_1}.
    \end{equation}
    In this sense, we do not need to deal with the different neutrino mass hierarchies separately anymore, and we can know the hierarchy through the sign of $\Delta$. 
    The positive $\Delta$ denotes the NH case, and the negative $\Delta$ denotes the IH case.
    Then, the three neutrino mass eigenstates can be expressed by $\Delta$
    \begin{align}
        &m_1=\frac{1-\Delta}{2} \sqrt{ \frac{\Delta m_{31}^2}{\Delta} },
        \label{eq:m1} \\
        &m_2 = \sqrt{\frac{(1-\Delta)^2}{4} \frac{\Delta m_{31}^2}{\Delta} +\Delta m_{21}^2},
        \label{eq:m2} \\
        &m_3 = \frac{1+\Delta}{2} \sqrt{ \frac{\Delta m_{31}^2}{\Delta} },
        \label{eq:m3}
    \end{align}
    and the total mass is given by
    \begin{equation}
        \sum m_\nu=\sqrt{\frac{\Delta m_{31}^2}{\Delta} } + \sqrt{\frac{(1-\Delta)^2}{4} \frac{\Delta m_{31}^2}{\Delta} +\Delta m_{21}^2} .
        \label{eq:smu}
    \end{equation}
    The total mass and the separate masses for different hierarchies are shown in Fig. \ref{fig:mass}. 
    We can see that the total mass is not less than $0.06$~eV for NH and $0.10$~eV for IH automatically.
    Besides, the current dimensionless energy density parameter for neutrinos is given by
    \begin{equation}
        \Omega_\nu = \frac{\sum m_\nu}{93.14 h^2 \text{eV} },
    \end{equation}
    where $h$ is the dimensionless Hubble constant expressed as $H_0 = 100 h \text{km s}^{-1} \text{Mpc}^{-1}$.
    
    Therefore, we can see that the hierarchy parameter $\Delta$, giving the two fixed mass splittings, is not only a description of the hierarchy, but also a way to code the lightest neutrino mass and the total neutrino mass.

\section{Methodology and data sets}
\label{sec:datmet}

\subsection{Methodology}
\label{sec:method}

    The publicly available \textbf{CosmoMC} code is used to perform Markov Chain Monte Carlo (MCMC) sampling in the present paper.
    The Boltzmann equation solver \textbf{CAMB} code \cite{Lewis:1999bs} is included in the \textbf{CosmoMC} for computation of cosmological quantities.
    According to Eqs. (\ref{eq:m1})-(\ref{eq:smu}) we modify the \textbf{CosmoMC}.
    The total neutrino mass and the corresponding lightest neutrino mass under different hierarchies can be derived by judging the sign of $\Delta$.
    Since an unknown parameter $\Delta$ is added, the corresponding parameter space for the $\Lambda$CDM model is
    \begin{equation}
        \{ \omega_b, \omega_c, 100 \theta_{MC}, \tau, n_s, \ln[10^{10} A_s], \Delta \},
        \label{eq:parameters}
    \end{equation}
    where $\omega_b = \Omega_b h^2$ and $\omega_c = \Omega_c h^2$ are the physical density of baryons and cold dark matter today, $\theta_{MC}$ is the ratio between the sound horizon and the angular diameter distance at the decoupling epoch, $\tau$ is the Thomson scattering optical depth, $n_s$ and $A_s$ are the spectral index and amplitude of scalar power spectrum, respectively.
    The extra free parameters for the $w$CDM model and the $w_0 w_a$CDM model are $w$ and $\{w_0, w_a\}$, respectively.

    Before considering the cosmological observational data, one is required to select prior probability distributions for the model's parameters.
    We use a flat linear prior for the cosmological parameters, they are: 
    $\omega_b  \in [0.005, 0.1], \omega_c \in [0.001, 0.99], 100\theta_{MC} \in [0.5, 10], \tau \in [0.01, 0.8], n_s \in [0.8, 1.2], \ln{[10^{10} A_s]} \in [2,4]$, and $w \in [-3, 0.5]$ for $w$ constant model while for CPL model $w_0 \in [-3, 0.5]$ and $w_a \in [-3, 3]$.
    We first apply a flat linear prior on $\Delta$ in the range $[-1,1]$ to do a standard MCMC sampling, then importance sample the chain with different priors on $\Delta$.
    Details are in Sec.~\ref{sec:importsamp}.
    
    The Bayesian evidence is a good measure of the statistical preference for a model over another one, by computing the Bayes factor \cite{Trotta:2008qt}.
    For a specific model $\mathcal{M}$, a set of parameters $\theta$, and the data sets $d$, the Bayesian evidence $E$, also called the marginal likelihood, is given by
    \begin{equation}
        E = p(d \mid \mathcal{M}) = \int_{\Omega_\mathcal{M}} p(d\mid \theta, \mathcal{M}) p(\theta\mid\mathcal{M}) d\theta,
        \label{eq:bayesE}
    \end{equation}
    where $p(\theta \mid \mathcal{M})$ is the prior on $\theta$ in model $\mathcal{M}$, $p(d\mid \theta, \mathcal{M})$ is the likelihood.
    Using the same data sets and assuming that different models have identical prior probabilities $p(\mathcal{M}_i)=p(\mathcal{M}_j)$, the posterior probability of the model $\mathcal{M}_i$ can be obtained by applying Bayes' theorem
    \begin{equation}
        p(\mathcal{M}_i\mid d) = \frac{p(\mathcal{M}_i) p(d\mid\mathcal{M}_i)}{p(d)} 
        \propto p(d\mid\mathcal{M}_i).
        \label{eq:Mpost}
    \end{equation}
    Thus, the Bayes factor $B_{ij}$ of model $\mathcal{M}_i$ with respect to model $\mathcal{M}_j$ is then given by
    \begin{equation}
        B_{ij}=\frac{p(\mathcal{M}_i\mid d)}{p(\mathcal{M}_j\mid d)}
        = \frac{E_i}{E_j} 
        ~~~ \Rightarrow ~~~
        \ln{B_{ij}} = \ln{E_i}-\ln{E_j}.
    \end{equation}
    The strength of the preference for one of the competing models over the other is usually determined by means of the Jeffreys scale \cite{Kass:1995loi, Trotta:2008qt}: $0< |\ln{B_{ij}}|<1$ is regarded as weak evidence, $1<|\ln{B_{ij}}|<3$ is positive, $3<|\ln{B_{ij}}|<5$ is strong, and $|\ln{B_{ij}}| > 5$ is very strong.

    Computing the Bayesian evidence is notoriously computationally expensive, however, progress has been made on this issue in \cite{Heavens:2017hkr, Heavens:2017afc}, where the $k$-th nearest neighbor distances are used to compute the Bayesian evidence from MCMC chains.
    The proposed algorithm is implemented in the \textbf{MCEvidence}\footnote{\url{https://github.com/yabebalFantaye/MCEvidence}}.
    In this work, we will use the code to compute the logarithm of the Bayes factor of different models and priors.

\subsection{Data sets}
\label{sec:datasets}

    Both dark energy and neutrinos can affect the CMB power spectrum by changing the acoustic peaks scale, the late ISW effect, etc
    \cite{Ade:2015rim}.
    The addition of the dynamical dark energy will increase the degeneracies of cosmological parameters, thus using the CMB power spectrum alone is not enough.
    To further constrain the properties of dark energy, we will combine some geometric observations at low redshifts, including the Baryon acoustic oscillations (BAO), type Ia supernova observation and direct measurements of the Hubble constant.
    These low redshift measurements can provide strong exploration on the EOS of dark energy at $z \lesssim 1$.
    The data combinations used in this article are listed as follows:
    \begin{itemize}
        \item the \textit{Planck} 2015 data release of CMB temperature and polarization anisotropies, including LowTEB, TT, EE, and TE. Details of the likelihood code can be found in \cite{Adam:2015rua,Aghanim:2015xee};
        \item the BAO data including the measurements from 6dFGS ($z_{\rm eff}=0.1$) \cite{Beutler:2011hx}, SDSS MGS ($z_{\rm eff}=0.15$) \cite{Ross:2014qpa}, CMASS ($z_{\rm eff}=0.57$) and LOWZ ($z_{\rm eff}=0.32$) samples of BOSS DR12  \cite{Gil-Marin:2015nqa}, and also the RSD data from the CMASS and LOWZ\footnote{Note that the BOSS DR12 BAO and RSD measurements are not considered at the same time. The BOSS DR12 results from BAO likelihood will not be included in our calculation when the RSD data are considered.} \cite{Gil-Marin:2015nqa};
        \item the Joint Light-curve Analysis (JLA) sample\cite{Betoule:2014frx}, compiled from the SNLS, SDSS, and the samples of several low-redshift of the SNe;
        \item the local measurement of $H_0$, i.e. Hubble Space Telescope (HST) 2016 with $H_0 = 73.03 \pm 1.79 {\rm  km s}^{-1} {\rm Mpc}^{-1}$ at $1\sigma$ confidence level (C.L.), which is in strong tension with CMB-only determinations  \cite{Riess:2016jrr}. 
    \end{itemize}
    Here observation of the BAO, JLA, and $H_0$ can partly break the degeneracies at the low redshifts, and the RSD data is used for constraining the neutrino masses well \cite{Zhao:2016ecj}.

\section{Results and Discussion}
\label{sec:result}

\subsection{Constraint on the neutrino masses and mass hierarchies}
\label{sec:results}

    Using the data sets mentioned in Section \ref{sec:datasets} and a modified version of \textbf{CosmoMC}, our constraints on the cosmological parameters of different cosmological models with flat linear prior on $\Delta$ are listed in Table \ref{tab:results}.
    In this table, the best-fit results are shown with the $68\%$ C.L. uncertainty of the cosmological parameters and $95\%$ C.L. upper limits of the neutrino mass hierarchy parameter $\Delta$, the neutrino's total mass $\sum m_\nu$, and the minimal mass $m_{\nu,{\rm min}}$.
    
\begin{figure*}
	\centering
	\includegraphics[scale=0.5]{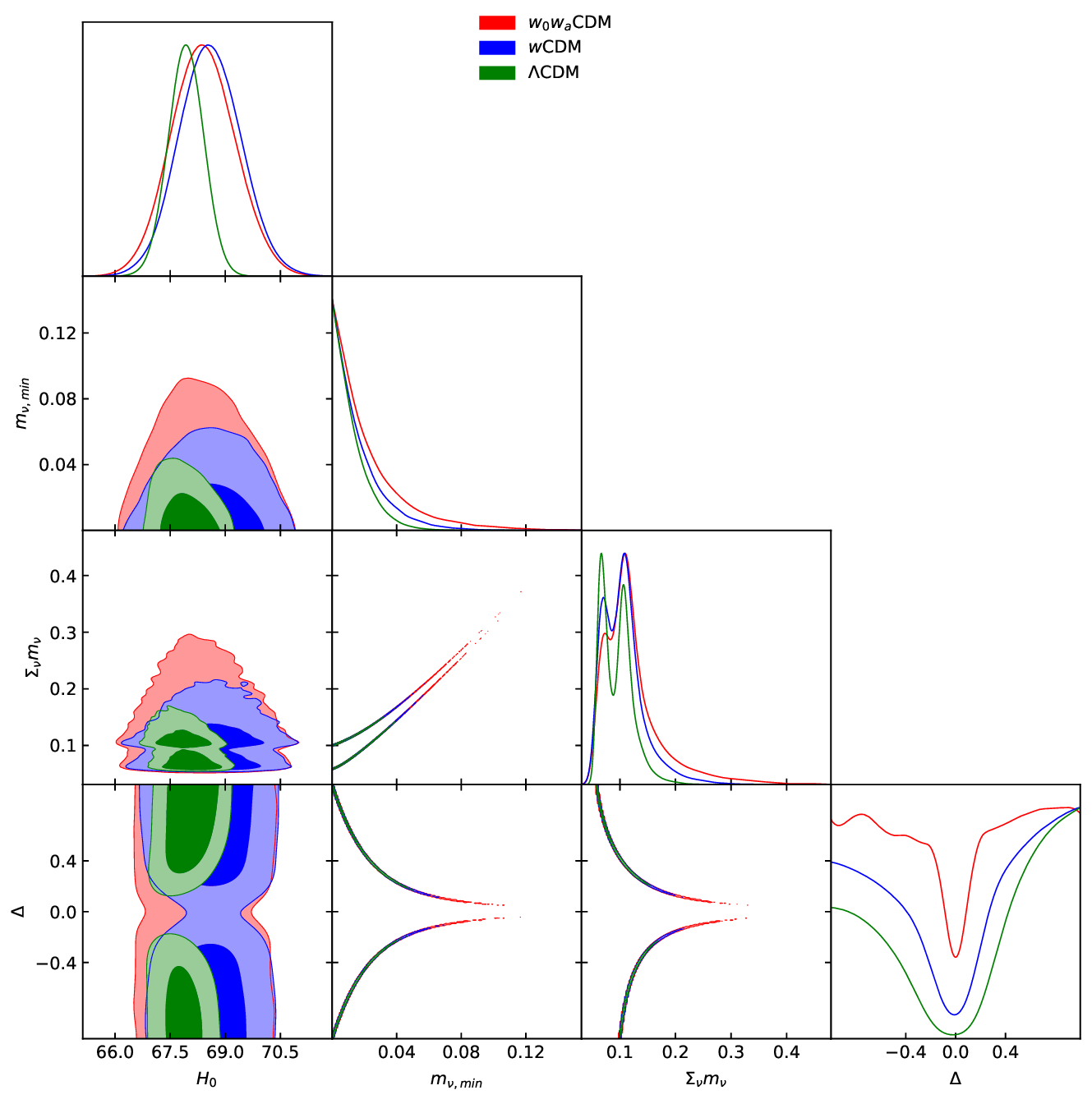}
	\caption{The 1D marginalized distribution and 2D contours for the interested parameters of different dark energy models.}
	\label{fig:results}
\end{figure*}

    Figure \ref{fig:results} shows the one-dimensional marginalized distribution and $68\%$, $95\%$ C.L. regions of some selected parameters for the three cosmological models.
    The figure shows that the 1D marginalized distribution of $\sum m_\nu$ has two peaks, which is quite different from the one presented in other published analysis.
    The bimodality of the marginalized distribution implies that there are two best-fit values of the parameter.
    So what causes the emergence of this?
    Here we will give an explanation.
    In our study, we use a hierarchy parameter to represent all neutrino masses and their total mass, so the total masses of NH and IH case are mixed together.
    Therefore, the probability distributions of $\sum m_\nu$ will roughly be $p(\Sigma m_\nu) = p(\Sigma^{\text{NH}} m_\nu) p(\text{NH}) + p(\Sigma^{\text{IH}} m_\nu) p(\text{IH})$.
    Obviously, their expectations are not necessarily the same and thus the distribution of $\sum m_\nu$ will have two peaks.
    From Sec. \ref{sec:delta}, we know that the minimal value of $\sum m_\nu$ for NH and IH are $0.06$~eV and $0.10$~eV, respectively,
    Thus, the first peak is the best-fit of $\sum^{\text{NH}}_\nu m_\nu$ and the second peak is the best-fit of $\sum^{\text{IH}}_\nu m_\nu$. 
    
    Furthermore, the higher the peak, the greater the probability.
    The height of the two peaks is different for the three cosmological models considered in this paper.
    The left peak is higher than the right peak in the $\Lambda$CDM model, the two peaks have the similar height in the $w$CDM model, and the left peak is lower than the right one in the $w_0 w_a$CDM model.
    However, it should be noted that the post-distribution is a mixed distribution of NH and IH case, so it's hard to say which hierarchy is favored.
    The post-distribution of $\Delta$ could provide more information about this issue.
    In Fig.~\ref{fig:results}, one can find that there are also two peaks in the posterior distribution of $\Delta$: one is NH ($\Delta >0$) and the other is IH ($\Delta<0$).
    The two hierarchies are no longer coupled together.
    Combined with the above analysis, we conclude that in the $\Lambda$CDM model, $\Delta > 0$ (NH) is favored over $\Delta < 0$ (IH), and $\Delta = 0$ is strongly disfavored.
    However, this trend is not so obvious in the $w$CDM model and even disappears in the $w_0 w_a$CDM model.
    
    In NH case, the lightest neutrino mass is less than $0.030$ eV, $0.039$ eV and $0.053$ eV for $\Lambda$CDM, $w$CDM and $w_0 w_a$CDM model at $95\%$ C.L., respectively.
    But in IH case, the corresponding values become $0.024$ eV, $0.035$ eV and $0.053$ eV.
    Further, the upper limit of $\sum m_\nu$ at $95\%$ C.L. of $\Lambda$CDM, $w$CDM, $w_0 w_a$CDM models are $0.119$ eV, $0.142$ eV, $0.179$ eV in the NH case, and $0.135$ eV, $0.158$ eV, $0.198$ eV in the IH case, respectively.
    From Eq. (\ref{eq:smu}), it can be deduced that the minimum value of $\sum m_\nu$ for NH and IH are $0.06$ eV and $0.10$ eV, respectively.
    Therefore, the upper limit of $\sum^{\text{NH}} m_\nu$ or $\sum^{\text{IH}} m_\nu$ in our constraints is not small enough to rule out the IH case.
    
    In general, the range of the hierarchy parameter $\Delta$ is significantly different in the $\Lambda$CDM, $w$CDM, and $w_0 w_a$CDM models.
    The $\Lambda$CDM model has the smallest range of $\Delta$, and the $w_0 w_a$CDM model has the largest one.
    According to the statistical analysis results of different parameters in Fig. \ref{fig:results} and Table \ref{tab:results}, it's easy to find that, compared with the $w$CDM model, the constraints on cosmological parameters are looser in the $w_0 w_a$CDM model and tighter in the $\Lambda$CDM model.

\begin{table*}
	\centering
	\label{tab:results}
	\renewcommand\arraystretch{1.2}
	\begin{tabular}{|cccc|}
		\hline
		Parameters  &  $\Lambda$CDM & $w$CDM & $w_0 w_a$CDM  \\
		\hline		
		$\Omega_b h^2$    & $0.02237\pm 0.00014$  & $0.02232\pm 0.00015$   & $0.02229\pm 0.00015$ \\
		$\Omega_c h^2$    & $0.1177\pm 0.0010$    & $ 0.1183\pm 0.0012$    & $0.1188\pm 0.0013$ \\
		$100\theta_{MC}$  & $1.04095\pm 0.00030$  & $1.04085\pm 0.00031$   & $1.04079\pm 0.00032$ \\
		$\tau$            & $0.078\pm 0.016$      & $0.073\pm 0.017$       & $0.070\pm 0.018$ \\
		$\ln[10^{10} A_s]$ & $3.086\pm 0.031$      & $3.076\pm 0.033$       & $3.071\pm 0.034$ \\
		$n_s$             & $0.9697\pm 0.0038$    & $0.9680\pm 0.0042$     & $ 0.9667\pm 0.0044$ \\		
		\hline
		$H_0$             & $67.93\pm 0.48$ & $68.59\pm 0.84$ & $68.40\pm 0.87$ \\
		$\Omega_{de}$     & $0.6942\pm 0.0062$ & $0.6985\pm 0.0077$ &  $0.6956\pm 0.0085$ \\
		$\Omega_m$        & $0.3058\pm 0.0062$ & $0.3015\pm 0.0077$ & $ 0.3044\pm 0.0085$ \\
		$\sigma_8$        & $0.815\pm 0.013$ & $0.821\pm 0.015$ & $0.820\pm 0.016$ \\
		$z_{\rm reion}$   & $9.9^{+1.5}_{-1.3}$ & $9.4^{+1.7}_{-1.4}$ & $ 9.1^{+1.7}_{-1.4}$ \\
		
		\hline
		$w/w_0$
		                  & $-1$ & $-1.035\pm 0.036$ & $-0.94\pm 0.11$ \\
		$w_a$
		                  & --- & --- & $-0.40^{+0.46}_{-0.34}$  \\
		
		\hline
		$\Delta$ $(95\%)$  & $[-1,-0.33) ~{\rm  or }~ (0.23,1]$ & $[-1,-0.20)~ {\rm  or } ~(0.16,1]$ & $[-1,-0.10)~ {\rm  or }~ (0.09,1]$ \\
		$m_{\nu,{\rm min}}^{\rm NH} {\rm eV } (95\%)$ 
		                   & $< 0.030$  & $< 0.039$ & $< 0.053$ \\
		$m_{\nu,{\rm min}}^{\rm IH} {\rm  eV } (95\%)$ 
		                   & $< 0.024$ & $< 0.035$ & $< 0.053$  \\
		$\sum^{\rm NH} m_\nu {\rm  eV } (95\%)$ 
		                   & $< 0.119$ &$<0.142$ &$< 0.179$ \\
		$\sum^{\rm IH} m_\nu {\rm  eV } (95\%)$ 
		                   & $< 0.135$ &$<0.158$ &$< 0.198$ \\
		\hline
	\end{tabular}
	\caption{Constraints on independent and derived cosmological parameters, EOS of dark energy models ($68\%$ C.L.), and $95\%$ upper limits for the total neutrino mass and the lightest neutrino.}
\end{table*}

\subsection{Massive neutrinos versus dark energy}

    As expected, the free parameters of the dark energy models have significant influences on constraining the cosmological parameters and the neutrino masses.
    The effects of dynamical dark energy on the cosmological measurements of neutrino masses under different cosmological observation data sets have been studied in \cite{Zhao:2016ecj}. 
    As a complement, we now focus on the correlation between neutrino hierarchy parameter and dark energy property.
    
\begin{figure}
	\centering
	\includegraphics[scale=0.5]{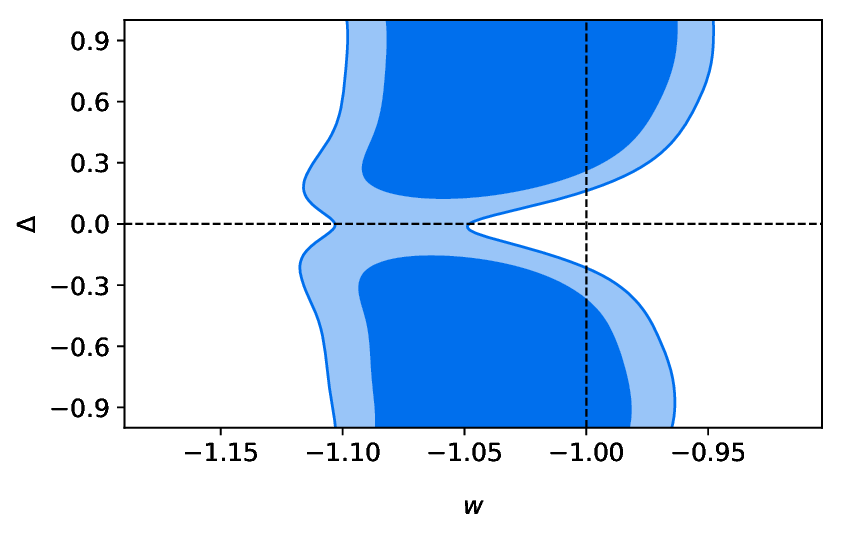}
	\caption{The $68\%$ and $95\%$ C.L. contours in the $w-\Delta$ plane of $w$CDM model.
	The $\Lambda$CDM case with $w=-1$ is shown in the planes by the dashed lines.}
	\label{fig:wcdm}
\end{figure}

\begin{figure}
	\centering
	\includegraphics[scale=0.5]{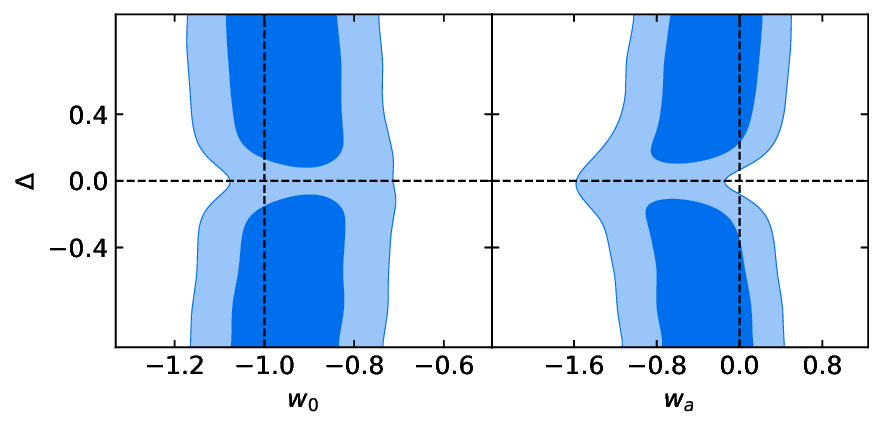}
	\caption{
	The $68\%$ and $95\%$ C.L. contours in the $w_0-\Delta$, $w_a-\Delta$ planes for the $w_0 w_a$CDM model.
	The vertical black dashed lines is $w_0=-1$ and $w_a=0$.}
	\label{fig:cpl} 
\end{figure}

    As Fig.~\ref{fig:wcdm} shows, in the $w$CDM model, NH ($\Delta >0$) is positively correlated with $w$, but IH ($\Delta < 0$) is negatively related to $w$.
    Note that $\sum m_\nu$ decreases as the absolute value of $\Delta$ increases. 
    Then it can be concluded that $w$ is anti-correlated with $\sum m_\nu$.
    Here, this correlation can be explained by the compensation for the effects on the acoustic peak scale $\theta_*$ \cite{Zhao:2016ecj}.
    For that, a reduction in $\sum m_\nu$ can compensate the changed $H(z)$ due to the increasing $w$.
    The contour also shows that for either NH case or IH case, a phantom like dark energy, i.e., $w<-1$ is favored.    
    For the $w_0 w_a$CDM model, in Fig.~\ref{fig:cpl}, we find that $w_0$ is positively correlated with $\sum m_\nu$ and $w_a$ is anti-correlated with $\sum m_\nu$.
    From the contours, it is clear that a dark energy model with time-varying EOS transforming from phantom like to quintessence like with cosmic expansion is favored by the current data.

    From Figs.~\ref{fig:wcdm} and \ref{fig:cpl}, one can also find that the contours in $w-\Delta$, $w_0-\Delta$, and $w_a-\Delta$ planes all looks symmetrical about $\Delta = 0$, which indicates that NH and IH are almost equally possible given a specific set of dark energy parameters.

\subsection{\boldmath Importance sampling and prior choices under different priors of $\Delta$}
\label{sec:importsamp}
    
    We have generated several MCMC chains with flat linear priors on $\Delta$, and now we wish to investigate the impacts of different $\Delta$ priors without running \textbf{CosmoMC} again and again, where the importance sampling technique \cite{Lewis:2002ah} will be adopted.
    In this case, importance sampling is independent of the cosmology, so a very simple procedure is possible and should work well: importance sampling the existing MCMC chains by multiplying the multiplicity of each point with different new priors \cite{Hannestad:2017ypp}.
    Three different priors will be considered, the flat logarithmic prior on the absolute value of the neutrino hierarchy parameter $\Delta$, the flat linear prior on the total neutrino mass $\sum m_\nu$, and the flat logarithmic prior on $\sum m_\nu$.
    The complete list of priors is listed in table \ref{tab:priors}.

\begin{figure}
    \centering
    \includegraphics[scale=0.5]{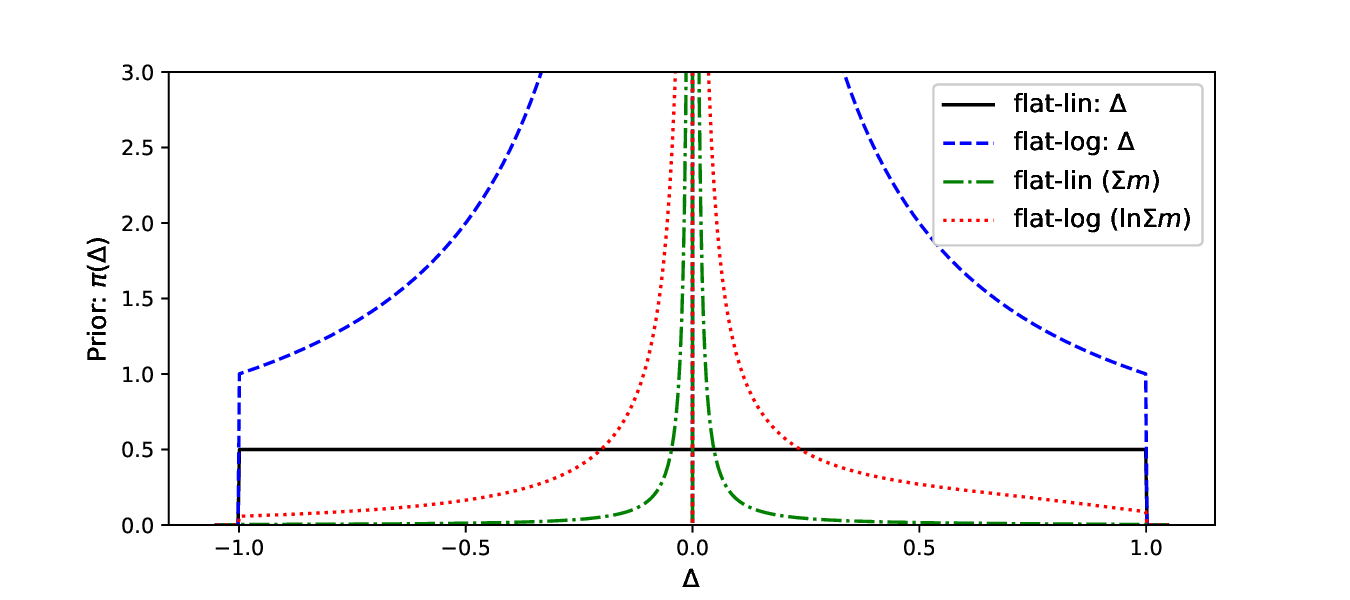}
    \caption{Different priors on $\Delta$.}
    \label{fig:priors}
\end{figure}

    \begin{table}
        \centering
        \begin{tabular}{|c|c|c|c|} 
            \hline
            Parameters & Prior & Range & Corresponding Prior on $\Delta$\\
            \hline
            $|\Delta|$      & logarithmic & $[10^{-4}, 1]$   & $\propto 1/|\Delta|$\\
            $\sum m_\nu$ eV & linear      & $[\Sigma, 7.50]$ & $\propto |d\Sigma m_\nu/d\Delta|$\\
            $\sum m_\nu$ eV & logarithmic & $[\Sigma, 7.50]$ & $\propto \frac{|d\Sigma m_\nu/d\Delta|}{\Sigma m_\nu}$\\
            \hline
        \end{tabular}
        \caption{The three new priors, where $\Sigma = 0.06$~eV and $0.10$~eV for NH and IH case, respectively.}
        \label{tab:priors}
    \end{table}
    The priors of $\Delta$ can be derived from the above definitions of different priors and Eq.~(\ref{eq:smu}).
    The different priors are shown in Fig.~\ref{fig:priors}.
    The three new priors increase the importance of a smaller $|\Delta|$, i.e., a larger $\sum m_\nu$.
    This effect will be shown in the posterior distributions, resulting in an increase in the upper limit of the total neutrino mass.

\begin{figure}
    \centering
    \includegraphics[scale=0.5]{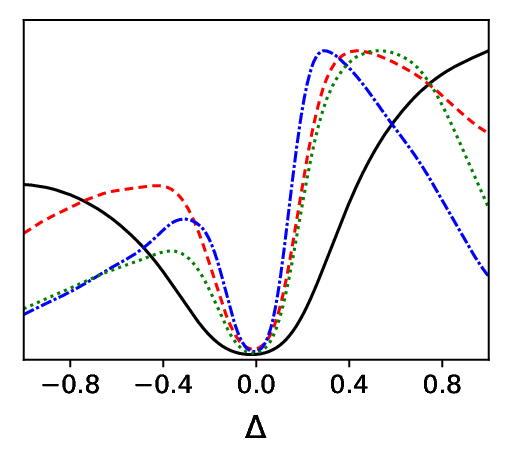}
    \includegraphics[scale=0.5]{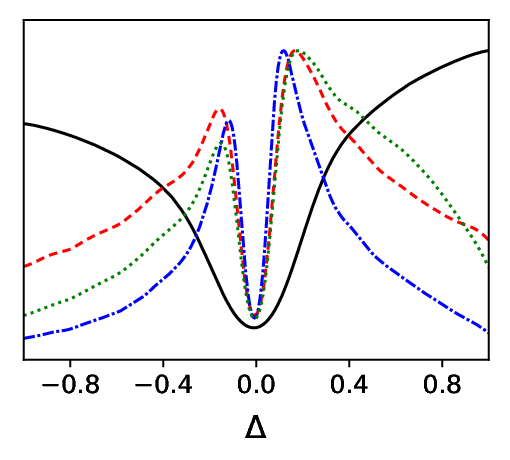}
    \includegraphics[scale=0.5]{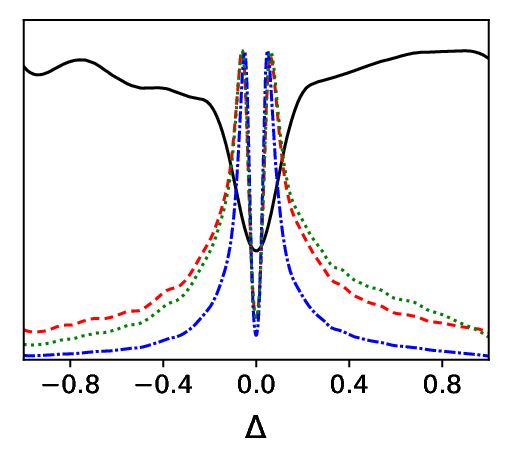}
    \caption{From left to right are for $\Lambda$CDM, $w$CDM and $w_0 w_a$CDM model, respectively.
    In the graphics, the black-solid line is for the flat-lin $\Delta$ case, red-dashed line is for the flat-log $\Delta$ case, blue-dotdashed line is for flat-lin $\sum m$ case, and green-dotted line is for flat-log $\sum m$ case. }
    \label{fig:posterior}
\end{figure}

    Our results of importance sampling are summarized in Table~\ref{tab:masses} and Fig.~\ref{fig:posterior}.
    One can find that the three new priors relaxed the constraints on $\sum m_\nu$.
    And the 2$\sigma$ upper limits of the total neutrino mass $\sum m_\nu$ can vary significantly from one prior to another.
    In contrast with Table~\ref{tab:results}, it is easy to find that the most strict constraints occur in the case that the prior on $\Delta$ is flat linear.
    The posterior distributions of $\Delta$ with different priors are shown in Fig.~\ref{fig:posterior}.
    Notice that the NH case is always favored in $\Lambda$CDM and $w$CDM model regardless of the four priors.
    Similarly to the case of linear prior on $\Delta$, neither the NH nor the IH case is favored in the $w_0 w_a$CDM model.

\begin{table}
    \centering
    \label{tab:masses}
    \renewcommand\arraystretch{1.2}
    \begin{tabular}{|cccccc|}
        \hline
        Prior &  Model & $m_{\nu, \text{min}}^\text{NH}$~eV & $m_{\mu,\text{min}}^\text{IH}$~eV & $\sum^{\text{NH}} m_\nu$~eV & $\sum^{\text{IH}} m_\nu$~eV  \\
        \hline
        flat-log $\Delta$ & $\Lambda$CDM & $< 0.0416$ & $< 0.0339$ & $< 0.149$ & $< 0.155$  \\
        flat-log $\Delta$ & $w$CDM       & $< 0.0654$ & $< 0.0602$ & $< 0.213$ & $< 0.217$  \\ 
        flat-log $\Delta$ & $w_0 w_a$CDM & $< 0.106$  & $< 0.106$  & $< 0.329$ & $< 0.341$  \\ 
        flat-lin $\sum m$ & $\Lambda$CDM & $< 0.0499$ & $< 0.0408$ & $< 0.171$ & $< 0.170$  \\ 
        flat-lin $\sum m$ & $w$CDM       & $< 0.0843$ & $< 0.0740$ & $< 0.267$ & $< 0.252$  \\
        flat-lin $\sum m$ & $w_0 w_a$CDM & $< 0.129$  & $< 0.125$  & $< 0.396$ & $< 0.393$  \\ 
        flat-log $\sum m$ & $\Lambda$CDM & $< 0.0423$ & $< 0.0319$ & $< 0.151$ & $< 0.150$  \\
        flat-log $\sum m$ & $w$CDM       & $< 0.0657$ & $< 0.0596$ & $< 0.214$ & $< 0.215$  \\ 
        flat-log $\sum m$ & $w_0 w_a$CDM & $< 0.106$  & $< 0.106$  & $< 0.330$ & $< 0.342$  \\ 
        \hline
    \end{tabular}
    \caption{Impacts of different priors on neutrinos masses at $95\%$ C.L.}
\end{table}

    Now, setting the $\Lambda$CDM model with flat linear prior on $\Delta$ as the base model, we conduct the Bayesian evidence analysis to compare the various cosmological models and prior choices.
    We have computed $\ln{B}$ to quantify the strength of the evidence for the base model.
    In particular, a negative value of $\ln{B_{ij}}$ ( $j$ is the base model, $i$ for other cases ) corresponds to evidence for the base model, and a positive value to evidence for the $i$-th model or case. 
    
    Table~\ref{tab:bayes} shows Bayes factors with respect to the base model.
    We see that under the same prior, there is a weak preference for the $\Lambda$CDM model over the $w$CDM model, but a very strong preference for the $w_0w_a$CDM model over the $\Lambda$CDM model.
    For the same cosmological model, the flat linear prior on $\Delta$ is always positively favored with respect to the logarithmic priors on $|\Delta|$ and $\sum m_\nu$ and strongly favored with respect to the linear prior on $\sum m_\nu$.
    Of all these cases, the $w_0 w_a$CDM model with flat linear prior on $\Delta$ is the most preferred one than other cases.
    In this subsection and the previous subsection, we have reached the conclusion that, in the $w_0w_a$CDM model, the cosmological observations used in this article do not show any preference for NH or IH.

\begin{table}
    \centering
    \label{tab:bayes}
    \renewcommand\arraystretch{1.2}
    \begin{tabular}{|cccc|}
         \hline
         Prior &  Model & $\ln{B}$ & $\Delta \ln{B_{ij}}$ \\
        \hline
        flat-lin $\Delta$ & $\Lambda$CDM &  -6872.02  &   0.00  \\
        flat-lin $\Delta$ & $w$CDM       &  -6871.48  &   0.54  \\
        flat-lin $\Delta$ & $w_0 w_a$CDM &  -6866.20  &   5.82  \\
        \hline
        flat-log $\Delta$ & $\Lambda$CDM &  -6874.36  &  -2.34  \\
        flat-log $\Delta$ & $w$CDM       &  -6873.59  &  -1.57  \\
        flat-log $\Delta$ & $w_0 w_a$CDM &  -6867.96  &   4.06  \\
        \hline
        flat-lin $\sum m$ & $\Lambda$CDM &  -6881.00  &  -8.98  \\
        flat-lin $\sum m$ & $w$CDM       &  -6880.01  &  -7.99  \\
        flat-lin $\sum m$ & $w_0 w_a$CDM &  -6874.04  &  -2.03  \\
        \hline
        flat-log $\sum m$ & $\Lambda$CDM &  -6874.34  &  -2.32  \\
        flat-log $\sum m$ & $w$CDM       &  -6873.60  &  -1.58  \\
        flat-log $\sum m$ & $w_0 w_a$CDM &  -6867.97  &   4.05  \\
        \hline
    \end{tabular}
    \caption{Values of the logarithm of Bayesian factor for different models with different priors with respect to $\Lambda$CDM model with flat linear prior on $\Delta$.}
\end{table}

\section{Conclusion} 
\label{sec:conclusion}

    Massive neutrinos have significant influences on the dynamics of the universe, such as the CMB anisotropy and the matter fluctuations.
    Therefore, the CMB power spectra and the large scale structure observations will provide potential methods of measuring the neutrino masses and the mass hierarchy, etc.
    However, this cosmological constraint could be affected by the properties of dark energy, due to the fact that dark energy also affects the CMB power spectra and the large scale structures.
    In order to find out the influence of dark energy on the neutrino mass hierarchy, we focus on two typical dynamical dark energy models, the $w$CDM and $w_0 w_a$CDM model.
        
    There are two different neutrino mass hierarchies, i.e., NH and IH.
    In order to reduce the number of computation workloads, we use the method proposed in \cite{Xu:2016ddc}, in which the sign of $\Delta$ is adopted to measure the neutrino mass hierarchy.
    In this way, once $\Delta$ is determined, it is easy to distinguish the NH ($\Delta >0$) from IH ($\Delta < 0$), and derive all of the three neutrino masses and the total neutrino mass.
    Using this method and the publicly available \textbf{CosmoMC} code, the models have been constrained using the  current observation data combination: \textit{Planck} 2015 LowTEB, TT, TE, EE + BAO DR12 + JLA SN + HST 2016.
    
    The MCMC samples results, with uniform priors on all the free parameters, are discussed in section \ref{sec:result}.
    By comparing the constrained results of the $w$CDM and $w_0 w_a$CDM models with the $\Lambda$CDM model, we find that, in addition to neutrino mass, the best-fit values of basic and derived cosmological parameters are almost identical for different models.
    For the neutrino mass, the tightest constraints occur in the $\Lambda$CDM model, and the loosest constraints occur in the $w_0 w_a$CDM model.
    According to the marginalized distribution of $\Delta$, we find that NH is preferred over IH in the $\Lambda$CDM and $w$CDM model, but it cannot distinguish NH from IH in the $w_0 w_a$CDM model.
    By analyzing the contours between $\Delta$ and the free parameters of EOS in the two dynamical dark energy models, we find that the effects of neutrino mass on the evolution of the universe have a partial degeneracy with dark energy.
    The total neutrino mass is anti-correlated with the EOS in the $w$CDM model, and larger $\sum m_\nu$ favors phantom dark energy.
    However, for the $w_0 w_a$CDM model, the situation is somewhat complicated due to the introduction of an additional free parameter.
    In Fig. \ref{fig:cpl}, it can be found that the total neutrino mass is positively correlated with the present EOS $w_0$, and is anti-correlated with the change rate of EOS $w_a$.
    Consequently, a larger $\sum m_\nu$ favors a larger current EOS and a greater decreasing rate, i.e., an early phantom like but late quintessence like dark energy model is favored by a larger $\sum m_\nu$.
    Moreover, their contour plots also show that for a given set of dark energy parameters, NH and IH are almost equally possible.
    This indicates that the NH and IH case have strong degeneracy.
    
    The priors of neutrino masses may have a dramatic impact on the posterior probability distributions, so three new priors (flat logarithmic prior on $|\Delta|$, flat linear prior on $\sum m_\nu$, and flat logarithmic prior on $\sum m_\nu$) have been investigated by adopting the importance sampling technique.
    We have found that the upper limits of $\sum^{\text{NH}} m_\nu$ and $\sum^{\text{IH}} m_\nu$ increases when the new priors are considered.
    However, these priors do not change the cosmological model's preference for NH or IH.
    Finally, we perform a model comparison analysis, in which we compute the Bayes factors of different cosmological models with different priors.
    The results are summarized in Table~\ref{tab:bayes}.
    We find that the linear prior on $\Delta$ is always preferred over the three other priors within the same cosmological model and $w_0 w_a$CDM is always preferred over $\Lambda$CDM and $w$CDM models within the same prior.
    And for all the cases we consider the $w_0 w_a$CDM model with linear prior on $\Delta$ is the most favored.
    In summary, our method shows that the cosmological data sets used in this paper cannot tell which type of neutrino mass hierarchy is the most preferred.

\section*{Acknowledgments}

    L.X is supported in part by National Natural Science Foundation of China under Grant No. 11275035, Grant No. 11675032 (People's Republic of China), and supported by ``the Fundamental Research Funds for the Central Universities" under Grant No. DUT16LK31.

\providecommand{\href}[2]{#2}\begingroup\raggedright\endgroup

\end{document}